\documentclass[12pt,preprint]{aastex6}

\usepackage{amsmath}
\usepackage{natbib}
\usepackage{xcolor}
\usepackage[]{graphics}
\usepackage{epsfig}
\usepackage{epstopdf}

\usepackage{mathrsfs}

\begin{document}

\title{Possibility of Identifying Matter around Rotating Black Hole with Black Hole Shadow}

\author{
   Zhaoyi Xu,\altaffilmark{1,2,3}
   Xian Hou,\altaffilmark{1,2,3}
  Jiancheng Wang \altaffilmark{1,2,3}
 }

\altaffiltext{1}{Yunnan Observatories, Chinese Academy of Sciences, 396 Yangfangwang, Guandu District, Kunming, 650216, P. R. China; {\tt zyxu88@ynao.ac.cn, xianhou.astro@gmail.com, jcwang@ynao.ac.cn
}}
\altaffiltext{2}{Key Laboratory for the Structure and Evolution of Celestial Objects, Chinese Academy of Sciences, 396 Yangfangwang, Guandu District, Kunming, 650216, P. R. China}
\altaffiltext{3}{Center for Astronomical Mega-Science, Chinese Academy of Sciences, 20A Datun Road, Chaoyang District, Beijing, 100012, P. R. China}

\shortauthors{Xu et al.}

\begin{abstract}
Shadows cast by the rotating black hole in perfect fluid matter are studied using analytical method. We consider three kinds of matter with standard equation of state in Cosmology: $-1/3$ (dark matter), 0 (dust) and $1/3$ (radiation). The apparent shape of the shadow depends on the black hole spin $a$ and the perfect fluid matter intensity $k$. We find that the shadow is a perfect circle in the non-rotating case ($a=0$) and a distorted one in the rotating case ($a\neq0$), similar to what is found for other black holes. Interestingly, different kinds of matter have different influence on the black hole shadow. Dark matter has the strongest effect, dust is the next and radiation is the weakest. By applying our result to the black hole Sgr A$^{*}$ at the center of the Milky Way, we find that the angular resolution required to distinguish among different kinds of matter around the black hole is not much higher than the resolution of current astronomical instruments would be able to achieve in the near future. We propose that observing the black hole shadow provides a possibility of identifying the dominant matter near the black hole.

\end{abstract}


\section{INTRODUCTION}
It is a consensus that supermassive black holes exist at the centres of most galaxies. One of the most attractive area in astrophysics is the direct detection of a black hole through imaging the silhouette (or shadow) of its event horizon and the study of the interaction between the black hole and its environment. Such study can provide us information on fundamental properties of the black hole and eventually test Einstein's theory of General Relativity (GR). The bright radio source Sgr A$^{*}$ at the center of the Milky Way is one of the most interesting black hole candidates \citep[e.g.,][]{2006ApJ...638L..21B}. Efforts on direct imaging the shadow of Sgr A$^{*}$ are ongoing using the sub-millimetre ``Event Horizon Telescope'' (EHT)\footnote{www.Eventhorizontelescope.org.} \citep{2008Natur.455...78D} based on the very-long baseline interferometry (VLBI).

\cite{1966MNRAS.131..463S} was the first to study the shadow of the Schwarzschild black hole. Since then, analytical studies of black hole shadows have been a hot topic in astrophysics, from the rotating regular Kerr black hole to various black holes in different space-time \citep[e.g.,][]{1966MNRAS.131..463S,1979A&A....75..228L,1973blho.conf..215B,2009PhRvD..80b4042H,2000CQGra..17..123D,2005PASJ...57..273T,2018PhRvD..97f4021T,2013JCAP...11..063W,2013Ap&SS.344..429A,2009IJMPD..18..983S,2012PhRvD..85f4019A,2013PhRvD..87d4057A,2012PhLB..711...10B,2013PhRvD..88f4004A,2017JCAP...10..051W,2010CQGra..27t5006B,2016PhRvD..94h4025Y,2017PhLB..768..373C,2016arXiv161009477D,2016PhRvD..93j4004A,2016PhRvD..94b4054A,2018arXiv180203276S,2014PhRvD..89l4004G,2018arXiv180404898P,2018EPJC...78...91E,2010PhRvD..81l4045A,2017arXiv171209793K,2017CaJPh..95.1299M,2018arXiv180104592V,2014PhRvD..90b4073P, 2015EPJC...75..399A,2017arXiv170709521A,2017arXiv170707125P,2018EPJC...78...91E,2015PhRvD..92h4005A,2015PhRvD..92j4031P,2015PhRvL.115u1102C,2018arXiv180203062G,2012PhRvD..86j3001Y,2018arXiv180503798C,2018GReGr..50...42C}. Particular attention has been paid to the black hole shadow of Sgr A$^{*}$ using both analytical method and numerical simulations, trying to, on the one hand, constrain the accretion models by comparing to the EHT observations of Sgr A$^{*}$ \citep[e.g.,][]{2000ApJ...528L..13F,2007CQGra..24S.259N,2010ApJ...717.1092D,2014A&A...570A...7M,2015ApJ...799....1C,2016ApJ...820..137B,2017ApJ...837..180G}, and on the other hand, test theories of gravity \citep[e.g.,][]{2006JPhCS..54..448B,2009PhRvD..79d3002B,2014ApJ...784....7B,2015ApJ...814..115P,2016PhRvL.116c1101J,2018NatAs.tmp...41M}. Recently, the black hole shadow in quintessence and dark matter halo have also been investigated by \cite{2017arXiv171102898P}, \cite{2017IJMPD..2650051A} and \cite{2018arXiv180408110H}, further enriching this research field.


One interesting question is, for galactic nuclei that have no obvious black hole activities, such as the Galactic center of the Milky Way, what kind of substances (such as dark matter, dust, or radiation) near the supermassive black hole is dominant? In principle, substances are characterized by their equation of state. Therefore, once the equation of state of the substance around the black hole is known, the substance category is determined, at least approximately. In addition, we expect that for different substances, their impact on the black hole shadow will be different. Inspired by this idea, we propose to determine the substance category around the black hole using the characteristics of the black hole shadow in the presence of the given substance.  Under the assumption of perfect fluid matter in Cosmology, the equations of state for different kinds of matter are, $-1/3$ (dark matter), 0 (dust) and $1/3$ (radiation). We can imagine that in reality, matter around the black hole could be a mix of the three kinds of substances. In order to simply the discussion, we note that the equation of state considered in this work represents only the dominant effect of various species forms near the black hole. That's to say, $-1/3$, 0 and $1/3$ mean that the dominant species near the black hole is, respectively, dark matter, dust and radiation. 

The paper is organized as follows. In Section \ref{metric}, we introduce the space-time metric for spherical symmetric and rotating black hole in perfect fluid matter. In Section \ref{motion}, we derive the complete null geodesic equations for a test particle in the rotating black hole space-time. In Section \ref{shadow}, we study the shadow cast by the rotating black hole in the presence of perfect fluid matter. In Section \ref{emission}, we calculate the energy emission rate of the black hole. We discuss our result by applying to the black hole Sgr A$^{*}$ and summarize in Section \ref{discussion}.

\section{BLACK HOLE SPACE-TIME IN PERFECT FLUID MATTER}
\label{metric}
\subsection{Spherical symmetric black hole in perfect fluid matter}
The spherical symmetric black hole space-time metric in perfect fluid matter is \citep{2003CQGra..20.1187K}
\begin{equation}
ds^{2}=-f(r)dt^{2}+\frac{dr^{2}}{f(r)}+r^{2}(d\theta^{2}+sin^{2}\theta d\phi^{2}),
\label{SBH1}
\end{equation}
with
\begin{equation}
f(r)=1-\dfrac{2M}{r}-\dfrac{k}{r^{3\tilde{\omega}+1}},
\label{FR}
\end{equation}
where $M$ is the black hole mass, $k$ is the intensity of the perfect fluid matter, and $\tilde{\omega}=p/\rho$ is the equation of state with $p$ the presser and $\rho$ the density.

\subsection{Rotating black hole in perfect fluid matter}
The rotating black hole space-time metric in perfect fluid matter is \citep{2017EPJP..132...98T}
\begin{equation}
ds^2=-\left(1-\dfrac{2Mr+k r^{1-3\tilde{\omega}}}{\Sigma^2} \right)dt^2    +\dfrac{\Sigma^2}{\Delta}dr^2 - \dfrac{2a\sin^2\theta(2Mr+k r^{1-3\tilde{\omega}})}{\Sigma^2}d\phi dt + \Sigma^2 d\theta^2 $$$$
+sin^2\theta \left(r^2+a^2+a^2sin^2\theta \dfrac{2Mr+k r^{1-3\tilde{\omega}}}{\Sigma^2}   \right)d\phi^{2},
\label{KBH1}
\end{equation}
where
\begin{align}
&\Sigma^{2}=r^{2}+a^{2}cos^{2}\theta,\\
&\Delta=r^{2}f(r)+a^{2},
\label{KBH2}
\end{align}
and $f(r)$ takes the same form as Eq. (\ref{FR}). 

In this work, we consider three different cases of $\tilde{\omega}$: $-1/3$ \citep[dark matter dominant,][]{2003CQGra..20.1187K,2017EPJP..132...98T,2015arXiv150804761M,2017PhLB..771..365H,2017arXiv171104542X}, 0 (dust dominant) and $1/3$ (radiation dominant).

\section{NULL GEODESICS}
\label{motion}
Following the general method outlined in the literature, we first study the geodesic structure of a test particle moving around the black hole in perfect fluid matter. We adopt the Hamilton-Jacobi equation and Carter constant separable method \citep{1968PhRv..174.1559C}. The most general form of the Hamilton-Jacobi equation reads as
\begin{equation}
\frac{\partial S}{\partial \sigma}=-\frac{1}{2}g^{\mu\nu}\frac{\partial S}{\partial {x^{\mu}}}\frac{\partial S}{\partial {x^{\nu}}}
\label{HJ1}
\end{equation}
with $S$ the Jacobi action and $\sigma$ an affine parameter along the geodesics. The separable solution of $S$ is
\begin{equation}
S=\frac{1}{2}m^2\sigma-Et+L\phi+S_r(r)+S_{\theta}(\theta)
\label{HJ2}
\end{equation}
where $m$, $E$ and $L$ are, respectively, the test particle's mass, energy and angular momentum, with respect to the rotation axis, while $S_r(r)$ and $S_{\theta}(\theta)$ are functions of $r$ and $\theta$, respectively. Combining Eq. (\ref{HJ1}) and Eq. (\ref{HJ2}), we get the full geodesic equations for a test particle around the rotating black hole in perfect fluid matter, which take the following forms
\begin{align}
\label{HJ3}
&\Sigma\frac{dt}{d\sigma}=\frac{r^2+a^2}{\Delta}[E(r^2+a^2)-aL]-a(aE\sin^2\theta-L),\\
&\Sigma\frac{dr}{d\sigma}=\sqrt{\mathcal{R}},\\
&\Sigma\frac{d\theta}{d\sigma}=\sqrt{\Theta},\\
\label{HJ4}
&\Sigma\frac{d\phi}{d\sigma}=\frac{a}{\Delta}[E(r^2+a^2)-aL]-\left(aE-\frac{L}{\sin^2\theta}\right),
\end{align}
where $\mathcal{R}(r)$ and $\Theta(\theta)$ read as
\begin{align}
\label{HJ5}
&\mathcal{R}(r)=[E(r^2+a^2)-aL]^2-\Delta[m^2r^2+(aE-L)^2+\mathcal{K}],\\
&\Theta(\theta)=\mathcal{K}-\left(  \dfrac{L^2}{\sin^2\theta}-a^2E^2  \right) \cos^2\theta,
\end{align}
with $\mathcal{K}$ the Carter constant. The dynamics of the test particle around the rotating black hole in perfect fluid matter are then fully described by these equations. Now we study the geometry of photons ($m=0$) near the black hole. The apparent shape of the black hole shadow is determined by the unstable circular orbit, which satisfies the condition
\begin{equation}
\mathcal{R}=\frac{\partial\mathcal{R}}{\partial r}=0.
\label{R1}
\end{equation}

For an observer at the infinity ($\theta=\pi/2$, photons arrive near the equatorial plane) and by introducing two impact parameters $\xi$ and $\eta$ as
\begin{equation}
\xi=L/E,  \;\; \;\; \;\;   \eta=\mathcal{K}/E^2,
\end{equation}
we have
\begin{align}
\label{R2}
&(r^2+a^2-a\xi)^2-[\eta+(\xi-a)^2](r^2f(r)+a^2)=0,\\
&4r(r^2+a^2-a\xi)-[\eta+(\xi-a)^2](2rf(r)+r^2f^{'}(r))=0.
\label{R3}
\end{align}
Combining Eqs. (\ref{R2}-\ref{R3}) results in
\begin{align}
&\xi=\frac{(r^2+a^2)(rf'(r)+2f(r))-4(r^2f(r)+a^2)}{a(rf'(r)+2f(r))},\\
&\eta=\frac{r^3[8a^2f^{'}(r)-r(rf'(r)-2f(r))^2]}{a^2(rf'(r)+2f(r))^2}.
\end{align}
Furthermore, we have
\begin{align}
\xi^2+\eta &=2r^2+a^2+\frac{16(r^2f(r)+a^2)}{(rf'(r)+2f(r))^2}-\frac{8(r^2f(r)+a^2)}{rf'(r)+2f(r)}\\
                &=2r^2+a^2+\frac{8\Delta[2-(rf'(r)+2f(r)]}{(rf'(r)+2f(r))^2}.
\end{align}
where
\begin{equation}
f'(r)=\dfrac{2M}{r^2}+\dfrac{k(3\tilde{\omega}+1)}{r^{3\tilde{\omega}+2}}.
\label{dFR}
\end{equation}
as can be calculated from Eq. (\ref{FR}).

\section{BLACK HOLE SHADOW}
\label{shadow}
Now we study the shadow of the rotating black hole in perfect fluid matter. First, as usually presented in the literature, we introduce the celestial coordinates $\alpha$ and $\beta$ as 
\begin{align}
& \alpha = \lim_{r_o\to \infty}\left( -r_o^2 \sin \theta_o \dfrac{d\phi}{dr}  \right),\\
& \beta = \lim_{r_o \to \infty}\left( r_o^2 \dfrac{d\theta}{dr}  \right).
\end{align}
Here we assume the observer is at infinity, $r_o$ is the distance between the black hole and the observer, and $\theta_o$ is the inclination angle between the rotation axis of the black hole and the line of sight of the observer.  $\alpha$ and $\beta$ are the apparent perpendicular distances of the shadow as seen from the axis of symmetry and from its projection on the equatorial plane, respectively. 

From the null geodesic equations (\ref{HJ3}-\ref{HJ4}), the relation between celestial coordinates and impact parameters can be expressed as
\begin{align}
& \alpha = -\dfrac{\xi}{sin \theta},\\
& \beta = \pm \sqrt{\eta + a^2\cos^2\theta -\xi^2\cot^2 \theta}.
\end{align}
In the equatorial plane ($\theta=\pi/2$), $\alpha$ and $\beta$ reduce to
\begin{align}
& \alpha = -\xi,\\
& \beta = \pm \sqrt{\eta }.
\end{align}

\begin{figure}[htbp]
  \centering
   \includegraphics[scale=0.35]{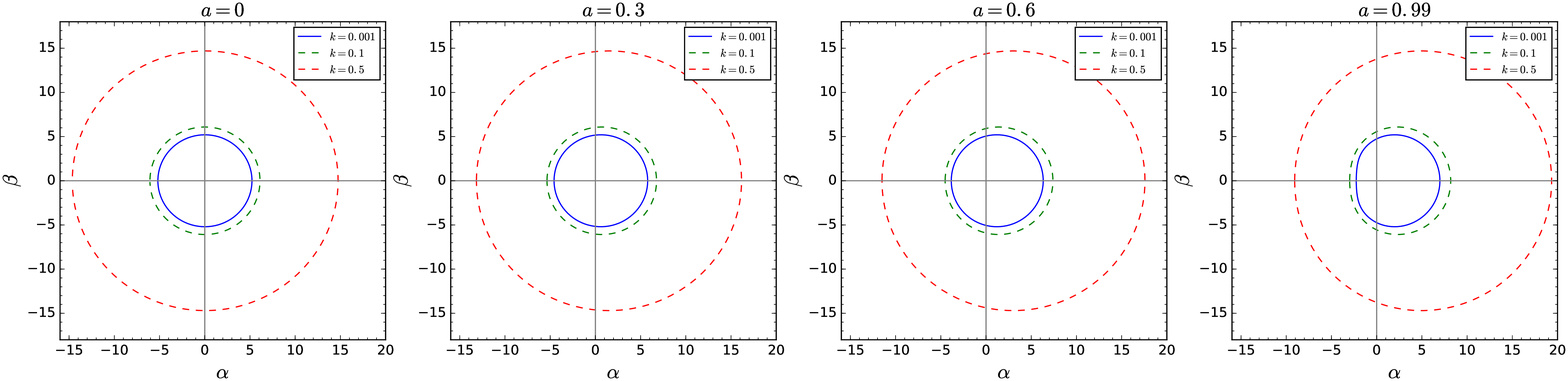}
   \includegraphics[scale=0.35]{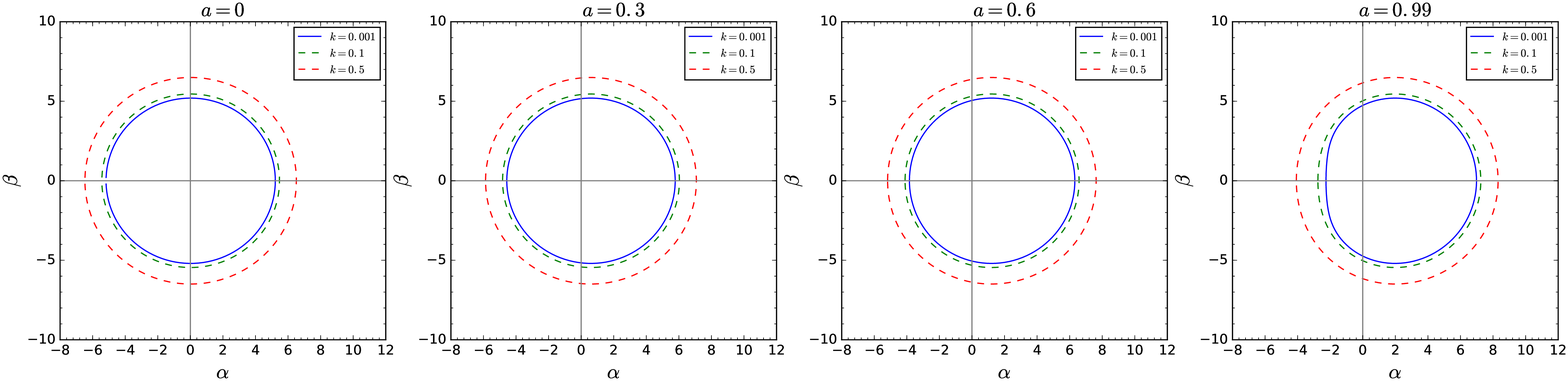}
   \includegraphics[scale=0.35]{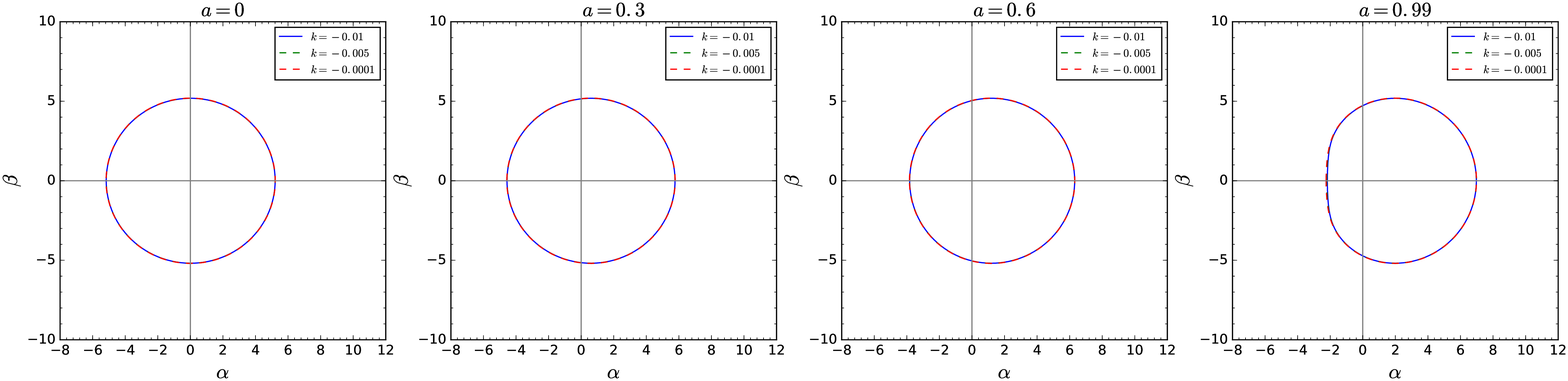}
   \caption{Silhouette of the shadow cast by the rotating black hole for different values of parameters $a$ and $k$, for the cases of dark matter ($\tilde{\omega}=-1/3$, upper panel), dust ($\tilde{\omega}=0$, middle panel) and radiation ($\tilde{\omega}=1/3$, lower panel).}
  \label{shadow_all}
\end{figure}

Different shapes of the shadow are shown in Figure \ref{shadow_all}. For dark matter ($\tilde{\omega}=-1/3$, upper panel) and dust ($\tilde{\omega}=0$, middle panel), it is clear that the shadow is a perfect circle and the size increases with the increasing $k$ in the non-rotating case ($a=0$). If considering the rotating case ($a\neq 0$), the shadow gets more and more distorted with the increasing $a$ and the size increases with the increasing $k$. Furthermore, when $k<0.1$, its effect on the shadow is minor even for a change of order of magnitude of 100 (from $k=0.001$ to $k=0.1$) for both the dark matter and dust cases, while as comparison, when $k>0.1$, an only 5 times larger value of $k$ leads already to a much more significant increase on the size of the shadow. In addition, the influence of $k$ in the case of dark matter looks much more significant than in the case of dust. For the case of radiation ($\tilde{\omega}=1/3$, lower panel), on the one hand, the shadow gets more distorted with the increasing $a$, similar to the cases of dark matter and dust; on the other hand, $k$ has incredible small even invisible effect on the size of the shadow. 

So, in general, different kinds of matter have different effects on the black hole shadow as we expected and dark matte has the strongest while radiation has the weakest. To explain this trend, we propose that as the perfect fluid pressure increases from dark matter to dust then to radiation, the mass density of the material will gradually decrease, so that the total mass of the material in the unit space under the same intensity (characterized by $k$) will decrease, and the gravitational force of the material on the black hole will also weaken. Therefore, it's natural that the impact on the black hole shadow follows the same pattern.

To discuss in detail the black hole shadow, we adopt the two astronomical observables defined in \cite{2009PhRvD..80b4042H}: the radius of the shadow $R_s$  and the distortion parameter $\delta_s$ \citep[c.f. Figure 3 in][]{2018arXiv180408110H}. $R_s$ is the radius of a reference circle passing through three points: $A(\alpha_r, 0)$, $B(\alpha_t, \beta_t)$ and $D(\alpha_b, \beta_b)$. $R_s$ approximately describes the size of the shadow and $\delta_s$ measures its deformation with respect to the reference circle. From the geometry of the shadow, we have
\begin{align}
& R_s = \dfrac{(\alpha_t-\alpha_r)^2 + \beta_t^2}{2|\alpha_r-\alpha_t|}, ~~~~~~~  \delta_s = \dfrac{d_s}{R_s} = \dfrac{|\alpha_p-\tilde{\alpha}_p|}{R_s},
\end{align}
where $d_s$ is the distance from the most left positions of the shadow $C(\alpha_p, 0)$ to the reference circle $F(\tilde{\alpha}_p, 0)$. 


It's clear that the shadow of the non-rotating ($a=0$) black hole is a perfect circle with radius of $R_s$. Thus
\begin{equation}
\alpha^2 + \beta^2 = \xi^2 +\eta= R_s^2. 
\end{equation}

\begin{figure}[htbp]
  \centering
   \includegraphics[scale=0.45]{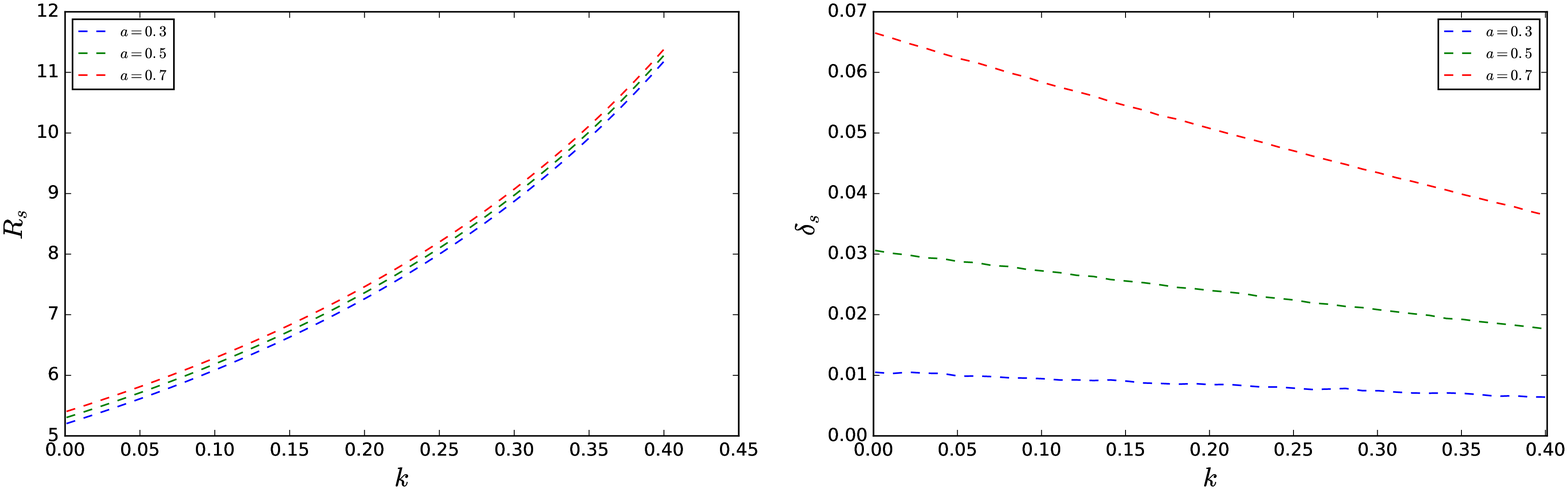}
   \caption{Variation of the radius $R_s$ (left) and the distortion parameter $\delta_s$ (right) of the black hole shadow  with the parameters $a$ and $k$ for the case of dark matter ($\tilde{\omega}=-1/3$). The lines of $R_s$ have been moved up vertically to visualize the trend of $R_s$ for different $a$ by adding a constant to $R_s$.}
  \label{Rs_omega1}
\end{figure}

\begin{figure}[htbp]
  \centering
   \includegraphics[scale=0.45]{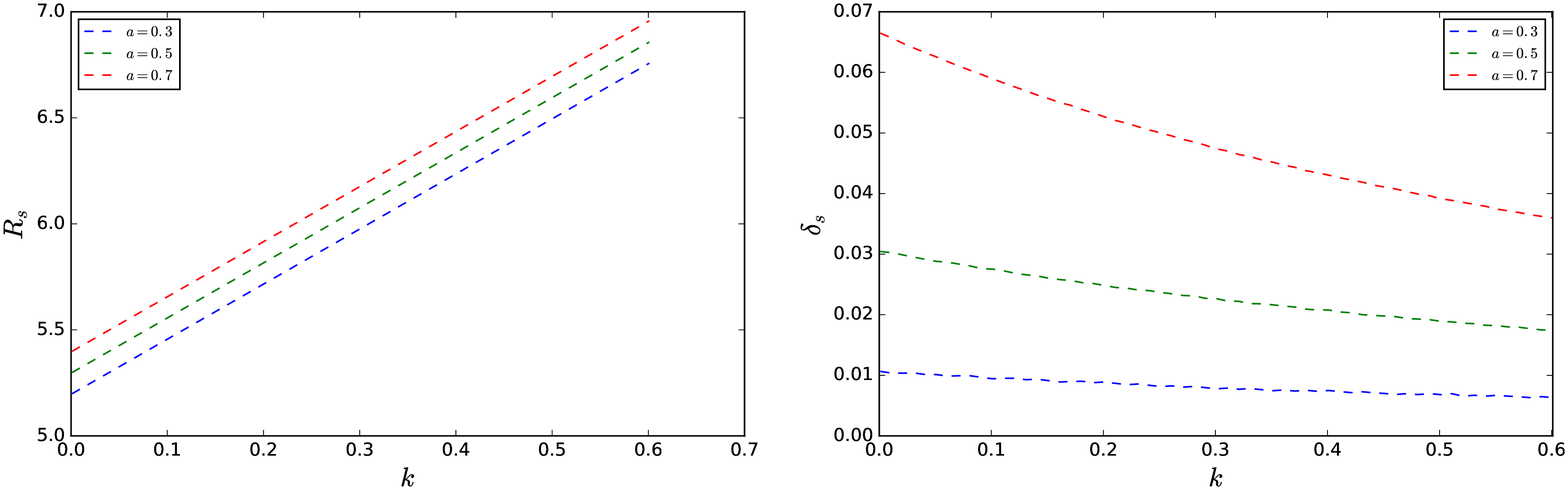}
   \caption{Variation of the radius $R_s$ (left) and the distortion parameter $\delta_s$ (right) of the black hole with the parameters $a$ and $k$ for the case of dust ($\tilde{\omega}=0$). The lines of $R_s$ have been moved up vertically to visualize the trend of $R_s$ for different $a$ by adding a constant to $R_s$.}
  \label{Rs_omega2}
\end{figure}

\begin{figure}[htbp]
  \centering
   \includegraphics[scale=0.45]{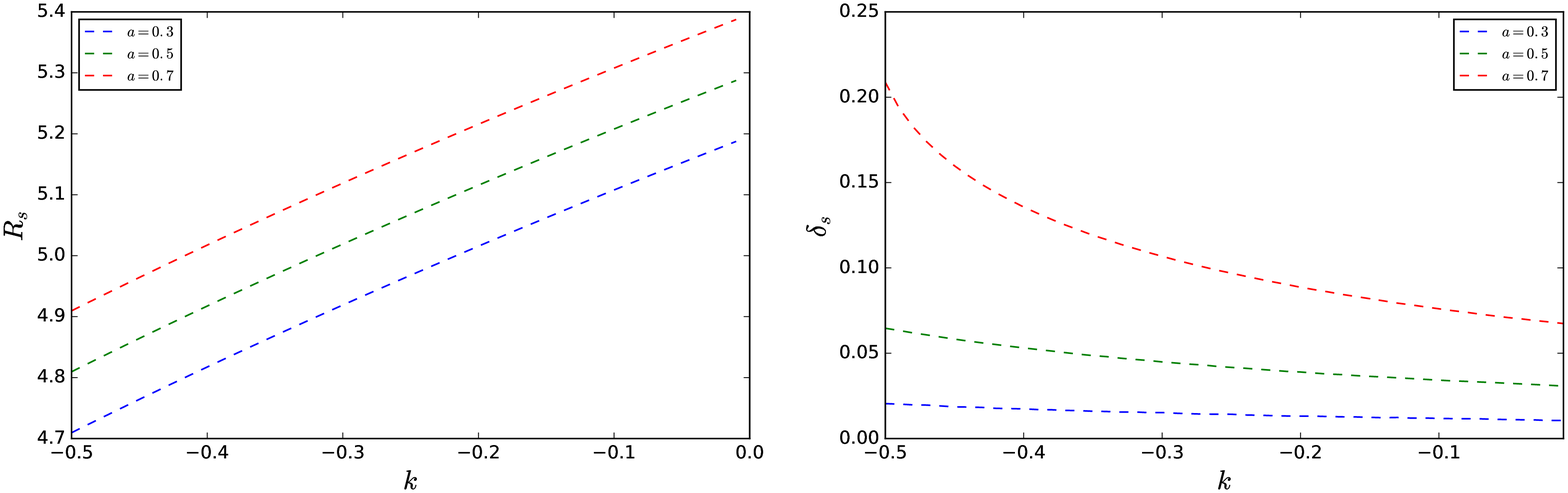}
   \caption{Variation of the radius $R_s$ (left) and the distortion parameter $\delta_s$ (right) of the black hole with the parameters $a$ and $k$ for the case of radiation ($\tilde{\omega}=1/3$). The lines of $R_s$ have been moved up vertically to visualize the trend of $R_s$ for different $a$ by adding a constant to $R_s$.}
  \label{Rs_omega3}
\end{figure}

Figure \ref{Rs_omega1}, \ref{Rs_omega2} and \ref{Rs_omega3} show the variation of the radius $R_s$ and the distortion parameter $\delta_s$ with the parameters $a$ and $k$ for the cases of $\tilde{\omega}=-1/3$ (dark matter), $\tilde{\omega}=0$ (dust) and $\tilde{\omega}=1/3$ (radiation), respectively. We find that the radius of the shadow increases monotonically with the increasing $k$, but almost does not vary with $a$ (a constant has been added to visualize the trend of $R_s$ with $a$). The distortion parameter decreases monotonically with the increasing $k$ for a given $a$, and increases with $a$ for a given $k$. This trend could also be inferred from Figure \ref{shadow_all}.

\section{ENERGY EMISSION RATE}
\label{emission}
As usually assumed in the literature, the black hole shadow corresponds to the high energy absorption cross section if the observer is located at an infinite distance. The high energy absorption cross section itself oscillates around a limiting constant value $\sigma_{lim}$.  For a spherical symmetric black hole, a good approximation is $\sigma_{lim}$ equals to the geometrical cross section of the photon sphere \citep{1973PhRvD...7.2807M,1973grav.book.....M} and can be calculated by \citep{2013JCAP...11..063W} 
\begin{equation}
\sigma_{lim} \approx \pi R_s^2,
\end{equation}
where $R_s$ is the radius of the black hole shadow. Given that the shadows discussed in our work approach to a standard circle (Figure \ref{shadow_all}), it's reasonable to apply this formula to the rotating black hole considered in this work. Thus, the energy emission rate of the black hole is
\begin{equation}
\dfrac{d^2E(\omega)}{d\omega dt} = \dfrac{2\pi^2\sigma_{lim} }{e^{\omega/T}-1}\omega^3,
\end{equation}
with $\omega$ the frequency of photon and $T$ the Hawking temperature for the outer event horizon ($r_+$). $T$ can be calculated from its definition
\begin{equation}
T = \lim_{\theta=0,  r \to r_+} \dfrac{\partial_r \sqrt{g_{tt}}   }{2\pi \sqrt{g_{rr}}}.
\end{equation}
In our case, we have (Eq. \ref{KBH1})
\begin{equation}
g_{tt} = 1-\dfrac{r^{2}-f(r)r^{2}}{\Sigma^{2}},   ~~~~~ g_{rr} = \dfrac{\Sigma^{2}}{\Delta}.
\end{equation}
Thus, the Hawking temperature is
\begin{equation}
T = \dfrac{r_+^2f'(r_+)(r_+^2+a^2) + 2a^2r_+(f(r_+)-1)  }{ 4\pi (r_+^2+a^2)^2      }.
\label{T_hawk}
\end{equation} 
If the perfect fluid matter is absent ($k=0$), Eq. (\ref{T_hawk}) reduces to the regular form for the Kerr black hole 
\begin{equation}
T_{Kerr} = \dfrac{r_+^2-a^2 }{ 4\pi r_+ (r_+^2+a^2) }
\end{equation} 
with $r_+ = M+\sqrt{M^2-a^2}$.

\begin{figure}[htbp]
  \centering
   \includegraphics[scale=0.45]{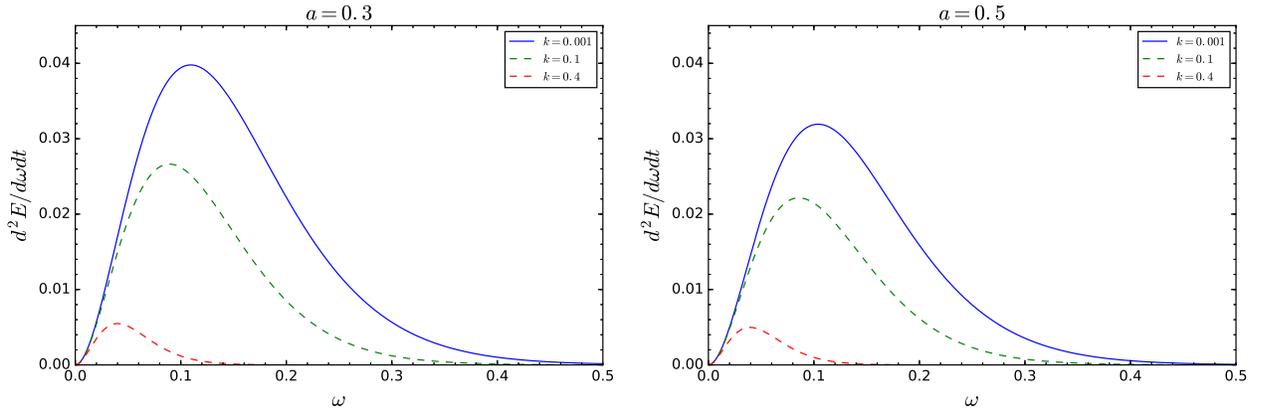}
   \caption{Evolution of the emission rate with the frequency $\omega$ for different values of the parameters $a$ and $k$ for the case of dark matter ($\tilde{\omega}=-1/3$).}
  \label{Erate_omega1}
\end{figure}

\begin{figure}[htbp]
  \centering
   \includegraphics[scale=0.45]{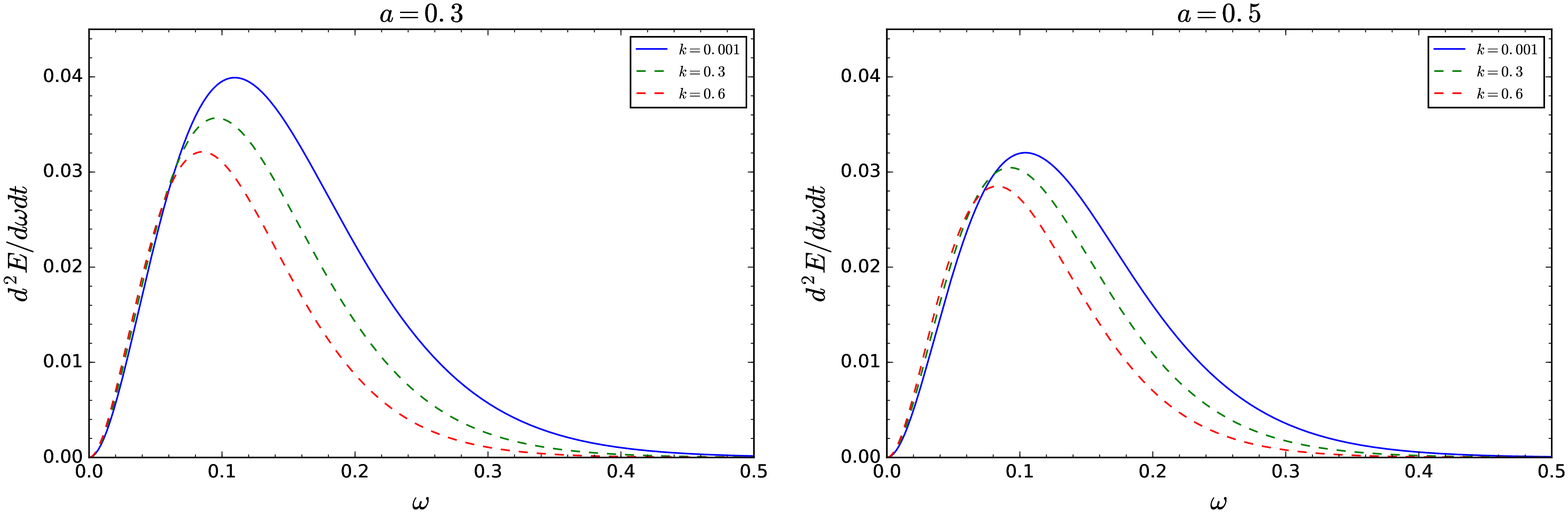}
   \caption{Evolution of the emission rate with the frequency $\omega$ for different values of the parameters $a$ and $k$ for the case of dust ($\tilde{\omega}=0$).}
  \label{Erate_omega2}
\end{figure}

\begin{figure}[htbp]
  \centering
   \includegraphics[scale=0.45]{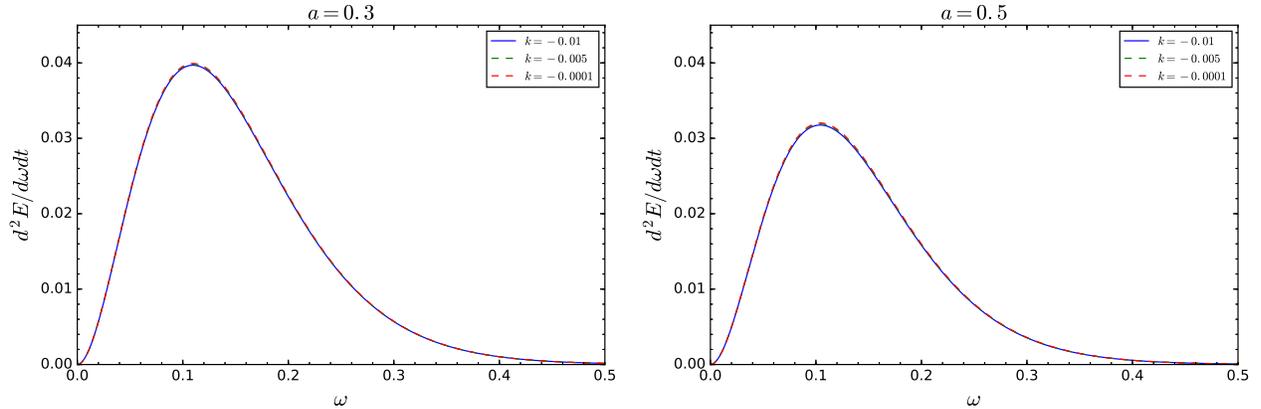}
   \caption{Evolution of the emission rate with the frequency $\omega$ for different values of the parameters $a$ and $k$ for the case of radiation ($\tilde{\omega}=1/3$).} 
  \label{Erate_omega3}
\end{figure}

Figure \ref{Erate_omega1}, \ref{Erate_omega2} and \ref{Erate_omega3} show the energy emission rate against the frequency $\omega$ for the cases of dark matter ($\tilde{\omega}=-1/3$), dust ($\tilde{\omega}=0$) and radiation ($\tilde{\omega}=1/3$), respectively. We can see that for the cases of dark matter and dust, the peak of the emission decreases with the increasing $k$ and shifts to lower frequency. For the case of radiation, the global shape of the emission rate is similar to the cases of dark matter and dust, but the variation of the emission rate with $k$ is tiny. This is similar to what we find for the evolution of shadow (Figure \ref{shadow_all}).

\section{DISCUSSION}
\label{discussion}
In this work, we investigate the shadow of the rotating black hole in three different perfect fluid matter backgrounds: dark matter, dust and radiation, represented by their equation of state $\omega$. For an observer located in the equatorial plane of the black hole, we show that the shadow is a perfect circle in the non-rotating case (spin $a=0$) and a deformed one in the rotating case ($a\neq 0 $). In general, the size of the shadow $R_s$ increases with the increasing $k$ which is the perfect fluid matter intensity parameter, and the deformation gets more and more significant (characterized by larger and larger distortion parameter $\delta_s$) with the increasing $a$. Furthermore, different kinds of matter influence the black hole shadow with different magnitudes. Dark matter has the most significant influence on the size of the shadow and dust has smaller one, while radiation has almost invisible impact even for a change of $k$ with an order of magnitude of 100.  In addition, assuming that the black hole shadow equals to the high energy absorption cross section, we calculate the emission rate of the black hole in the three different kinds of perfect fluid matter. We find that for the cases of dark matter and dust,  the emission rate decreases with the increasing $k$ and the peak of the emission shifts to lower frequency $\omega$. For the case of radiation, however, the emission rate almost does not vary with $k$, which is similar to the evolution of shadow. 
 
From the observational point of view, it is necessary to calculate the angular radius of the shadow which is related to the angular resolution of astronomical instruments. The angular radius can be estimated as $\theta_s = 9.87098\times 10^{-6} R_s(M/M\odot)(1 \rm{kpc}/$ $D)$ $\,\mu$as \citep{2012PhRvD..85f4019A} where $M$ is the black hole mass and $D$ is the distance from the black hole to the observer. For the supermassive black hole Sgr A$^{*}$ at the center of the Milky Way, $M=4.3\times10^6 M\odot$ and $D=8.3$ kpc. Thus, the observables $R_s$ and $\delta_s$ as well as the angular radius $\theta_s$ can be calculated if assuming an observer located in the equatorial plane of the black hole. We summarize the result in Table \ref{tab_thetas}. We find that: (1) For the three different cases of $\omega$ (dark matter, dust and radiation), the angular radius increases with the increasing $k$, which implies that the angular resolution required to detect the effect of perfect fluid matter on the black hole shadow decreases with $k$. (2) For the cases of $\omega=-1/3$ (dark matter) and $\omega=0$ (dust), the angular resolution needed to distinguish one from the other decreases with the increasing $k$. As examples, for $k=0.001$, the resolution would be 0.1 $\mu$as, for $k=0.1$, the resolution would be 1 $\mu$as and for $k=0.5$, the resolution becomes 10 $\mu$as. The last one is not much beyond the future resolution of EHT (15 $\mu$as at 345 GHz) and the space-based VLBI RadioAstron \citep[$\sim 1-10$ $\mu$as,][]{2013ARep...57..153K} \footnote{http://www.asc.rssi.ru/radioastron/index.html}. The less than 1 $\mu$as resolution required with smaller $k$ would be achieved by future instruments with highly improved techniques. (3) For the case of $\omega=1/3$ (radiation), the variations of both the observables and the angular radius with $k$ are much less significant than for the cases of dark matter and dust. This is consistent with what can be inferred from the evolution of the black hole shadow and the emission rate.

We summarize that different kinds of matter would have very different effects on the shadow of the black hole. This implies that it would be possible to identify the dominant substance around the black hole when first images of the black hole (e.g., Sgr A$^{*}$) are successfully obtained using the current and future astronomical instruments. This would eventually shed light on the nature of black holes and deepen our understanding of the black hole physics in general.

\setlength{\tabcolsep}{3pt}
\begin{table*}
\centering 
\caption{\small The observables $R_s$, $\delta_s$ and the angular radius $\theta_s$ for the supermassive black hole Sgr A$^*$ at the center of the Milky Way, for different cases of equation of state $\omega$, assuming a black hole spin of $a=0.3$.}
\begin{tabular}{lccccccccccccccc}
\toprule %
 $\omega$ & &   \multicolumn{3}{c}{$-1/3$ (dark matter)} & & &\multicolumn{3}{c}{0 (dust)}  & & & \multicolumn{3}{c}{$1/3$  (radiation) }  \\
\hline
 $k$ & &0.001  &0.1  &0.5  & & &0.001  &0.1  &0.5   & & &$-0.01$  &$-0.005$  &$-0.0001$\\
\hline
$R_s$ & &  5.204 & 6.086  &14.697  & & &5.199  &5.456  &6.495   & & &5.192  &5.19607  &5.19615  \\
$\delta_s $ (\%) & &1.05 & 0.94 &0.52  & & &1.06 &0.98 &0.68       & & &1.076 &1.073 & 1.067 \\
$\theta_s$ & &26.61  &31.12  &75.16   & & &26.59  &27.90   &33.22   & & &26.55  &26.5721  &26.5725  \\
\hline

\end{tabular}
\flushleft

\label{tab_thetas}
\end{table*}

\acknowledgments
We acknowledge the anonymous referee for a constructive report that has significantly improved this paper. We acknowledge the financial support from the National Natural Science Foundation of China under grants No. 11503078, 11573060 and 11661161010.

\bibliography{shadow_BH_matter}

\end{document}